\documentstyle[epsfig,12pt,tighten,aps]{revtex}
\begin{document}

\begin{titlepage}

\rightline{{\tt August 2002}}

\vskip 1.8 cm

\centerline{{\Large \bf Cosmological implications of low scale}}
\vskip 0.16cm \centerline{{\Large \bf quark-lepton unification}}

\vskip 1.3 cm

\centerline{\large T. L. Yoon and R. Foot}

\vskip 1.0 cm
\noindent

\centerline{{\it School of Physics}}
\centerline{{\it Research Centre for High Energy Physics}}
\centerline{{\it The University of Melbourne}}
\centerline{{\it Victoria 3010 Australia }}

\vskip 2.0cm

\centerline{Abstract} \vskip 0.7cm \noindent There is a unique
$\hbox{SU(4)}\otimes\hbox{SU(2)}_L \otimes \hbox{SU(2)}_R$ gauge
model which allows quarks and leptons to be unified at the TeV
scale -- thereby making the model testable and avoiding the gauge
hierarchy problem. In its minimal form, this model could quite
naturally accommodate simultaneous solutions to the solar and LSND
neutrino oscillation data. The atmospheric neutrino anomaly can be
easily accommodated by mirror-symmetrising the minimal model. The
model also contains three right-handed neutrinos, with masses in
the range 1 keV to $\sim$ 1 GeV. We investigate the implications
of these right-handed neutrinos for early Universe cosmology. It
is shown that the minimal model is inconsistent with some of the
standard assumptions of the big bang model. This motivates an
examination of non-standard big bang cosmology, such as a low
reheating temperature scenario with $T_{RH} \sim$ MeV. In such a
Universe, peaceful co-existence between low-scale quark-lepton
gauge unification and early Universe cosmology is possible.

\end{titlepage}

\section{Introduction \label{Intro}}

The similarities between the quarks and leptons does hint at the
possibility of a symmetry between them. The first example of such
a theory is the Pati-Salam model\cite{ps}, which is based on the
$\hbox{SU(4)}\otimes\hbox{SU(2)}_L \otimes \hbox{SU(2)}_R$ gauge
group. While a very good idea, this particular model has a number
of serious drawbacks. One problem is that experiments constrain
the high symmetry breaking scale to be greater than $20$ TeV
\cite{willen}, making direct tests of the model impossible, at
least within the next quarter century. Moreover, the presence of a
high symmetry breaking scale, significantly greater than a TeV,
becomes theoretically problematic since it leads to the gauge
hierarchy problem. This is quite unlike the situation in the
standard model where there is no `hierarchy problem' as such
because there is only one scale in the Higgs potential\footnote{It
is sometimes asserted in the literature that a hierarchy problem
also exists between the weak scale and the `Planck' scale; however
in this case things are much less clear. Firstly, the physics of
the Planck scale is really not understood at the moment so it is
not yet clear whether this (as yet unknown) physics will lead to a
fine-tuning problem. Another line of argument asserts that there
exists a fine-tuning problem between the weak scale and a momentum
`cut-off'. However that argument depends on the regulator scheme
used \cite{veltman}, [see also Ref. \cite{popovic}] and should not
be taken too seriously.}. However in extensions of the standard
model involving two (or more) symmetry breaking scales, there can
be a significant fine- tuning problem if the scales are widely
separated and appear in the Higgs potential. In that case, the fine-tuning
problem can be alleviated if one also assumes that low energy
supersymmetry exists. Unfortunately, this does not completely
solve the problem, though, since one still needs to arrange the
hierarchy at tree level. Also, low energy supersymmetry generates
a host of new problems such as sparticle mediated FCNC, rapid
proton decay, $\mu$-problem, etc., so that it ends up creating
more problems than it solves -- clearly an unsatisfactory
situation.

Perhaps an interesting question is the following one: Is it
possible to build a simple gauge model which unifies quarks and
leptons at low scales ($\stackrel{<}{\sim}$ few TeV) so that the
gauge hierarchy problem is avoided? The first such model was
written down some time ago \cite{QLS}, whereby a leptonic
SU(3)$_{\ell}$ gauge group was introduced allowing for a discrete
quark-lepton (spontaneously broken) symmetry to exist. The main
problem with the discrete symmetry approach comes from neutrino
masses. The lightness of the neutrino masses in that model (as
with the usual Pati-Salam model) suggests a high symmetry breaking
scale ($\gg$ TeV) if the usual see-saw mechanism is employed, and
while there are alternatives \cite{Foot:1995}, they are somewhat
complicated. Searches for new ideas led to the alternative
$\hbox{SU(4)}\otimes\hbox{SU(2)}_L \otimes \hbox{SU(2)}_R$ gauge
model\cite{422.1} which allows for TeV scale quark-lepton
unification without any problems for existing experiments with
{\it necessarily} tiny neutrino masses. This `alternative 422
model' predicts a multitude of new phenomenology including: rare
$B, K$ decays, baryon number violation as well as non-zero
neutrino masses, all of which are within current bounds, despite
the low symmetry breaking scale of a few TeV [see Refs.
\cite{422.2,422.3} for more details of the phenomenological
implications of the model].

In Ref. \cite{422.3} the extent to which the minimal alternative
422 model could accommodate solutions to the neutrino physics
anomalies was investigated. While it did not seem possible to
simultaneously accommodate solutions to all three classes of
neutrino physics anomalies [i.e. LSND \cite{LSND}, solar
\cite{solarNsk,solarsk,SNOnew} and atmospheric \cite{atmNsk,atmsk}
neutrino anomalies], it was shown that the minimal alternative 422
model can quite naturally accommodate the LSND and atmospheric
neutrino anomalies. The solar neutrino problem could be explained
if the model was extended with a mirror sector \cite{422.3,EPM}.
Since that time, there has been an important new development. SNO
has now measured both neutral- and charged-current solar neutrino
fluxes, providing strong evidence that large angle active-active
oscillations are occurring \cite{SNOnew}. In view of this new
development, the available information suggests an essentially
unique picture \cite{snopuz}:
\begin{eqnarray}
\nu_e &\to &\nu_\tau \ {\rm \ large \ angle \ oscillations \ explains \
the \ solar
\ neutrino \ problem}
\nonumber \\
\nu_\mu &\to &\nu_s \ {\rm \ large \ angle \ oscillations \ explains \
the \ atmospheric \ neutrino
\ anomaly}
\nonumber \\
\bar \nu_e &\to &\bar \nu_\mu \ {\rm \ small\ angle \ oscillations \ explain
\ LSND \ data}
\label{unique}
\end{eqnarray}
where $\nu_s$ is a hypothetical effectively sterile neutrino.
While the atmospheric neutrino data prefer the $\nu_\mu
\rightarrow \nu_{\tau}$ channel, it is also true that the $\nu_\mu
\rightarrow \nu_{s}$ possibility is only mildly disfavoured,
at the $\sim 1.5 \sigma- 3\sigma$ level, depending on how the data
are analysed \cite{atmsktau,Fexclude}.
The overall goodness of fit (g.o.f) of the above scheme has recently
been explicitly calculated in Ref.~\cite{footoct}, where it was found to be
0.26. That is, there is a $26\%$
probability of obtaining a worse global fit to the neutrino data.
This shows that the above scheme still provides a
reasonable fit to the totality of the neutrino oscillation
data\footnote{Recently, Ref.~\cite{Maltoni:2002xd} has
argued that all 4-neutrino models of
the (2~+~2) variety are ``ruled out'' by the totality of
neutrino oscillation data (solar, atmospheric and LSND).
However the g.o.f. obtained by Ref.~\cite{Maltoni:2002xd}
(g.o.f. = $10^{-6}$)
is not really the g.o.f but some other quantity. Indeed, as
shown in Ref.~\cite{footoct}, it disagrees with the actual
g.o.f by more than 5 orders of magnitude.
For more discussion on this issue, see
Ref.~\cite{footoct}.}.
Although this scheme is not
particularly popular, it at least has the
virtue that it will be tested in the near future:
MiniBooNE will test the oscillation explanation of the LSND
anomaly, while the forthcoming long baseline experiments
(MINOS and CNGS) will discriminate between the $\nu_\mu
\rightarrow \nu_{sterile}$ and $\nu_\mu \rightarrow \nu_{\tau}$
channels used to resolve the atmospheric neutrino anomaly.

The oscillation scheme [Eq.~(\ref{unique})] is essentially
unique in the sense that it is the simplest scheme
involving only two-flavour oscillations
explaining the totality of the data and also specific
features such as SNO's neutral current/charge current
solar flux measurement \cite{SNOnew}.
Of course, other, but more
complicated schemes involving multi-flavor oscillations, are
possible because they can also provide an acceptable fit to the data
for a range of parameters. For example, one can have
an additional parameter, $\sin^2 \omega$, where $\sin^2 \omega = 0$
corresponds to the scheme, Eq.~(\ref{unique}),
$\sin^2 \omega = 1$ is similar to Eq.~(\ref{unique}) with $\nu_\tau$
interchanged with $\nu_s$ and intermediate values of $\sin^2 \omega$
corresponds to mixed active + sterile oscillations \cite{other}.
Such schemes are called 2 + 2 models
because they feature two pairs of almost degenerate states
separated by the LSND mass gap.
While the scheme of Eq.~(\ref{unique}) could be viewed
as a particular 2 + 2 scheme with $\sin^2 \omega = 0$, it could
alternatively be viewed as motivating the
following hypothesis \cite{snopuz}:
{\em The fundamental theory of neutrino mixing, whatever it is,
features (i) large (or even maximal) $\nu_\mu \to \nu_s$
mixing, (ii) small-angle active-active mixing except for the
$\nu_e \to \nu_\tau$ channel which is large.}

This hypothesis is the one which we adopt in this paper.
We shall show that the minimal alternative
422 model is a candidate for the new physics required to explain
the active-active oscillations suggested by the LSND and solar neutrino
data within the above hypothesis. The atmospheric neutrino
anomaly, as explained above, will be assumed to be due
to $\nu_\mu \to \nu_s$ oscillations. The 422 model
does not have any suitable candidates for the needed
light sterile neutrino (it does have effectively
sterile right-handed neutrinos, but it turns out
that they are too heavy \cite{422.3}).

It is known \cite{EPM} that three light effectively
sterile neutrinos ($\nu'_e, \nu'_\mu, \nu'_\tau$)
maximally mixed with their active partners
is predicted to exist if mirror symmetry is an exact fundamental symmetry.
Thus, if we mirror symmetrise the model, we can easily
accommodate the atmospheric neutrino anomaly, via
$\nu_\mu \to \nu'_\mu$ oscillations.
If the oscillation lengths of the $\nu_{e,\tau} \to \nu'_{e,\tau}$
oscillations are longer than the Earth-Sun distance for
solar neutrinos, then this 3 active + 3 mirror neutrino
model effectively reduces to the required four-neutrino
scheme, Eq.~(\ref{unique}). It is also possible to
have the oscillation lengths of the $\nu_{e,\tau}\to \nu'_{e,\tau}$
oscillations shorter than the Earth-Sun distance. This
would mean a large sterile component ($\sim 50\%$) in the solar
neutrino flux, which is still allowed by the data \cite{bar}.
[It would also require some modifications to the solar
model, such as larger boron flux etc].

It turns out that the
consistency between the low symmetry breaking scale and the
solutions to these neutrino anomalies imposes some constraints on
the possible forms that could be assumed by the Majorana mass
matrix of the right-handed neutrinos (which form part of the
particle content of the model). As a result the masses of the
right-handed neutrinos in the alternative 422 model are
constrained to be in the ranges of 1 keV - 10 keV and 4 MeV to
$\sim$ 1 GeV [see the forthcoming Eq. (\ref{range}) discussed in
Section II].

These particles can potentially contribute significantly to the
matter density of the Universe. Within the framework of the
standard big bang model of cosmology, an important constraint is
that a given particle species $X$ must satisfy the cosmological
energy density bound
\begin{equation}\label{cosmic}
\Omega_{X} \equiv {\rho_X / \rho_c} \stackrel{<}{\sim} 1,
\end{equation}
where $\Omega_X$ is the contribution of their present density
$\rho_X$ normalised to the critical density, $\rho_c = 10^4 h^2
\hbox{ eV} \hbox{cm}^{-3}$ ($h =
H_0/100\hbox{kms}^{-1}\hbox{Mpc}^{-1}$ is the normalised Hubble
constant). It has been a routine practice to check if new particle
species contained in a given extension of the standard model are
compatible with standard cosmology and other astrophysical bounds.
This `consistency check' will form the content of the first half
of the present paper. It is found that even the lightest
right-handed neutrinos contained in this model are {\it not}
consistent with standard big bang cosmology.

Despite the claims that cosmology has ushered into an era of
unprecedented precision, there are still many unsettled issues at
the interface of particle physics and cosmology [see, for example,
Ref. \cite{albrecht} for a general account of pending issues
facing particle cosmology]. The inconsistency of a particle
physics model with standard cosmology does not necessarily mean
that a given particle physics model is not realistic. The standard
big bang cosmology is generally not robust against plausible
modifications (either by new observations or theoretical input) to
a set of standard assumptions. Hence it requires some cautious
attitude when one is to use standard cosmology to `rule out' or
support a given extension of the standard model of particle
physics. For example, the standard cosmological model contains an
implicit but not observationally justified assumption that the
reheating temperature (corresponding to the highest temperature
during the radiation-dominated epoch) of the early Universe
is much higher than
the characteristic temperatures of cosmological processes under
investigation (e.g. the freeze-out temperature pertinent to a
given particle species). However, Refs.
\cite{Kawasaki:1999na,Kawasaki:2000en} show that the reheating
temperature that is consistent with the observational light
element abundances could be as low as 0.7 MeV. The possibility of
a low reheating temperature scenario has prompted many works since
then (see for example Refs. \cite{Giudice,Giudice2,liuchun,kudo}).
For instance, Refs. \cite{Giudice,Giudice2} have shown that in a
low reheating cosmology the relationship between the relic density
of an exotic particle species normalised to the present energy
density of the Universe deviates from that of the standard case.
As a result, well-known constraints (such as the Cowsik-McClelland
bound \cite{CW}) previously imposed on the masses of the ordinary
neutrinos can be greatly relaxed in such a scenario.

In the low reheating temperature scenario, we shall re-analyse the
cosmological constraints imposed on the right-handed neutrinos
contained in the alternative 422 model. We find that a low
reheating temperature cosmology is consistent with the alternative
422 model, as there is some parameter space in which the
right-handed neutrinos can accommodate the cosmological and other
astrophysical bounds. We also point out that the lightest
right-handed neutrinos of the model provide an interesting dark
matter candidate.

The plan of this paper is as follows. In Section II we first
briefly explain how the minimal version of the alternative 422
model accommodates the solutions to the LSND and solar neutrino
data, which consistency requires the mass spectrum of the
right-handed neutrinos to fall in a specific range. Then, in
Section III we examine the cosmological implications of these
right-handed neutrinos in the framework of standard cosmology.
Having shown that the right-handed neutrinos are inconsistent with
standard cosmology, we proceed (in Section IV)  to examine the
right-handed neutrinos in the non-standard low reheating
temperature scenario with $T_{RH}\ {\sim}$ a few MeV. Taking the
reheating temperature $T_{RH}$ as a free parameter with a rough
lower bound of $T_{RH} \stackrel{>}{\sim}$ 0.7 MeV, we show that
the right-handed neutrinos could circumvent the cosmological
energy density bound for some parameter range. In Section V we
conclude.

\section{The masses of the right-handed neutrinos in the
alternative 422 model}

We first revise the details of the model and explain how it can
accommodate the large angle $\nu_e \rightarrow \nu_\tau$ and small
angle ${\nu_\mu} \rightarrow{\nu_e}$ oscillations suggested by the
solar and LSND anomalies. We refer the reader to Ref. \cite{422.3}
for further details.

The gauge symmetry of the alternative 422 model is
$\hbox{SU(4)}\otimes\hbox{SU(2)}_L \otimes \hbox{SU(2)}_R$. Under
this gauge symmetry the fermions of each generation transform in
the anomaly-free representations:
\begin{equation}\label{fermion}
Q_L \sim (4,2,1),\  Q_R \sim (4, 1, 2), \ f_L \sim (1,2,2).
\end{equation}
The minimal choice of scalar multiplets which can both break the
gauge symmetry correctly and give all of the charged fermions mass
is
\begin{equation}
\chi_L \sim (4, 2, 1), \ \chi_R \sim (4, 1, 2),\ \phi \sim
(1,2,2). \label{3}
\end{equation}
These scalars couple to the fermions as follows:
\begin{equation}
{\cal L} = \lambda_1 \hbox {Tr} \bigg[ \overline{Q_L} (f_L)^c
\tau_2 \chi_R\bigg] + \lambda_2 \hbox {Tr} \bigg[\overline {Q_R}
f^T_L \tau_2 \chi_L \bigg]+ \lambda_3 \hbox {Tr} \bigg[\overline
{Q_L} \phi \tau_2 Q_R\bigg]  + \lambda_4 \hbox {Tr}
\bigg[\overline {Q_L} \phi^c \tau_2 Q_R\bigg] + \hbox{H.c.},
\label{L1}
\end{equation}
where the generation index has been suppressed and $\phi^c =
\tau_2 \phi^* \tau_2$. The model reduces to the standard model
following the spontaneous symmetry breaking pattern:
\begin{eqnarray}
&\hbox{SU(4)}\otimes  \hbox{SU(2})_L \otimes \hbox{SU(2)}_R&
 \nonumber \\
&\downarrow \langle \chi_R \rangle& \nonumber \\ &\hbox{SU(3)}_c
\otimes \hbox{SU(2)}_L \otimes \hbox{U(1)}_Y & \nonumber \\
&\downarrow \langle \phi \rangle, \langle \chi_L \rangle \nonumber
\\ &\hbox{SU(3)}_c \otimes \hbox{U(1)}_Q&.
\end{eqnarray}
Note that the SU(4) group has a maximal SU(3)$_c \otimes $
U(1)$_T$ subgroup with the $4$ representation having the branching
rule $4 = 3(1/3) + 1(-1)$. The vacuum expectation values (VEVs)
can be conveniently expressed in terms of the $T, I_{3R}, I_{3L}$
charges as
\begin{eqnarray}
\langle \chi_R (T = -1, I_{3R} = 1/2) \rangle = w_R, \ \langle
\chi_L (T = -1, I_{3L} = 1/2) \rangle = w_L, \nonumber \\ \langle
\phi (I_{3L} = -I_{3R} = -1/2)\rangle = u_1,\ \langle \phi (I_{3L}
= -I_{3R} = 1/2)\rangle = u_2.
\end{eqnarray}
Note that $Y = T +2I_{3R}$ is the linear combination of $T$ and
$I_{3R}$ which annihilates $\langle \chi_R \rangle$ (i.e.
$Y\langle \chi_R \rangle = 0$) and $Q = I_{3L} + {Y \over 2}$ is
the generator of the unbroken electromagnetic gauge symmetry.
Observe that in the limit where $w_R \gg w_L, u_1, u_2$, the model
reduces to the standard model at low energies.

The identity of the particles in the fermion multiplets, Eq.
(\ref{fermion}), can now be made explicit. We have the known
quarks and leptons, along with some exotic heavy leptons $\{E^0,
E^{-}\}$ and a right-handed neutrino, $\tilde \nu_R$:
\begin{eqnarray}
Q_L^{\alpha,\gamma} = \left(\begin{array}{cccc} \tilde U & {\tilde
E}^0
\\ D & \tilde E^- \end{array}\right)_L,
Q_R^{\beta,\gamma} = \left(\begin{array}{cc} \tilde U & \tilde \nu
\\ \tilde D & \em l
\\
\end{array}
\right)_R, \ f_L^{\alpha, \beta} = \left(\begin{array}{cc} (\tilde
E^-_R)^c & \tilde \nu_L \\ (\tilde E^0_R)^c & \em l_L
\end{array}
\right). \label{multi}
\end{eqnarray}
In the above representation, $\alpha =\pm {1 \over 2} $, $\beta
=\pm {1 \over 2}$ index the SU(2)$_L$ and SU(2)$_R$ components
respectively. The SU(4) decomposition into the SU(3)$_c \otimes
$U(1)$_T$ maximal subgroup is indexed by $\gamma = \gamma'({1
\over 3}) \oplus 4(-1)$, where $\gamma' = y,g,b$ is the usual
colour index for SU(3)$_c$ and $\gamma' =4$ refers to the fourth
colour. The number in the bracket refers to the $T$ charge of the
subgroup U(1)$_T$. In the above matrices the first row of $Q_L$
and $f_L \ (Q_R)$ is the $I_{3L} \ (I_{3R}) = \alpha \ (\beta)=
1/2$ component while the second row is the $I_{3L} \ (I_{3R}) =
\alpha \ (\beta) = -1/2$ component. The columns of $Q_L, Q_R$ are
the $\gamma'({1 \over 3})$ and $4(-1)$ components of SU(4), and
the columns of $f_L$ are the $I_{3R} = \beta = \pm 1/2$
components. Each field in the multiplets Eq. (\ref{multi})
represents $3\times 1$ column vector of three generations. The
tilde in the fermion fields signify that they are the weak
eigenstates, which are generally not aligned with the
corresponding mass eigenstates.

A set of theoretically arbitrary CKM-type unitary matrices are
introduced into the theory to relate the weak eigenstates with
their corresponding mass eigenstates. The basis is chosen such
that
\begin{eqnarray}
E_R &=& V^\dagger {\tilde E}_R, \;\;\ E_L = U^\dagger {\tilde
E}_L, \nonumber \\ U_R &=& Y_R^\dagger{\tilde U}_R,\;\;\ U_L =
Y_L^\dagger{\tilde U}_L, \nonumber \\ D_R &=& K'{\tilde D}_R,
\label{basis}
\end{eqnarray}
and $\tilde D_L = D_L, \tilde {\em l}_L = {\em l}_L, \tilde {\em
l}_R = {\em l}_R$. The matrix $Y_L^\dagger \equiv K_L$ is the
usual CKM matrix (as in the standard model), whereas
$Y_R^\dagger{K'}^\dagger \equiv K_R$ is the analogue of the
CKM-type matrix for the right-handed charged quarks in the
SU(2)$_R$ sector. The matrix $K'$ is the analogue of the CKM-type
matrix in the SU(4) sector pertaining to lepto-quark interactions.

The model contains new gauge bosons: $W'$, $W_R^\pm$ and $Z'$. The
masses of these new bosons and the exotic leptons $\{E^0, E^{-}\}$
are constrained to be within the range:
\begin{eqnarray}\label{Mbound}
    0.5 (1.0) \hbox{ TeV} &\stackrel{<}{\sim}& M_{W_R},M_{Z'} (M_{W'})
    \stackrel{<}{\sim} 10 \hbox{ TeV}, \nonumber \\
    45 \hbox{ GeV} &\stackrel{<}{\sim}& M_{E_i}
    \stackrel{<}{\sim} 10 \hbox{ TeV}.
\end{eqnarray}
Note that the lower limit on the mass of the $E$ leptons arises
from LEP measurements of the $Z^0$ width, whereas the lower bound
on the masses of $Z', W_R$ is obtained from the consistency of the
model with LEP data \cite{422.1}. The upper bound is a rough
theoretical limit $-$ the scale of symmetry breaking cannot be
much greater than a few TeV, otherwise we would have a gauge
hierarchy problem. By this we mean a real fine-tuning problem in
the Higgs potential, not a hypothetical problem with the Plank
scale or artificial cut-off parameter.

At tree level, mixing between $\tilde \nu_R$ with $E_{L,R}^0$ in
the Lagrangian density of Eq. (\ref{L1}) generates the $3 \times
3$ Majorana mass matrix $M_R$ for the right-handed neutrinos after
spontaneous symmetry breaking (SSB):
\begin{eqnarray}\label{MR}
M_R &\simeq& \left(M_{\em l}V M_E^{-1} U^\dagger Y_L M_u
Y_R^\dagger \right) + \left(M_{\em l}V M_E^{-1} U^\dagger Y_L M_u
Y_R^\dagger\right)^\dagger,
\end{eqnarray}
where $M_{\em l}, M_u,  M_E$ are the $3 \times 3$ diagonal mass
matrices for the standard charged leptons, up-type quarks and the
exotic $E$ leptons. We know that the CKM matrix is approximately
diagonal, $Y_L \approx {\bf I}_3$, whereas the rest of the
CKM-type matrices (which are not present in the standard model)
are poorly constrained by experiments (except for the matrix $K'$
which is constrained to only a few specific forms by limits on
rare $K^0$ and $B^0$ decays mediated by the $W'$ bosons, see Ref.
\cite{422.2}). In the special case of decoupled generations (e.g.
$Y_L = Y_R = U = V = \bf I$), $M_R$ reduces to a diagonal matrix
$M_R = \hbox{ diag }\{2 m_{u} m_e / M_{E_1}, 2 m_{c} m_\mu /
M_{E_2}, 2 m_{t} m_\tau / M_{E_3} \}$.

At the one-loop level the gauge interactions from the charged
SU(2)$_L$ gauge bosons $W_L^{\pm}$ and SU(2)$_R$ gauge bosons
$W_R^{\pm}$ give rise to $\overline {\tilde \nu_L}(\tilde\nu_L)^c$
Majorana mass $m_M$, $\overline {\tilde \nu_L} \tilde \nu_R$ Dirac
mass $m_D $ and $\overline {\tilde \nu_L}(E^0_L)^c$ mass mixing
term $m_{\nu E}$\footnote{There is also radiative neutrino masses
arising from the scalar sector of the model. But because of the
larger arbitrariness in the scalar interactions we do not consider
them in depth.}. The Majorana mass $m_M$ is generically very small
in comparison to $m_D$ and $m_{\nu E}$, and can be ignored in the
subsequent discussion. The Dirac mass matrix $m_D$ is generated
[Fig. 1] via the diagonal (i.e. no cross generation mixing) gauge
interactions
\begin{equation}\label{glgr}
{g_L \over \sqrt 2} \overline {\tilde \nu_L} \not \! \!
W^+_{L}{\em l}_L + {g_R \over \sqrt 2} \overline {\tilde \nu}_R
\not \!\! W^+_{R}{\em l}_R + \hbox{ H.c.},
\end{equation}
where $g_L$ is the usual SU(2)$_L$ coupling constant, and $g_R$ is
the SU(2)$_R$ coupling constant. As discussed in Ref.
\cite{422.1}, $g_R(M_{W'}) \approx g_L(M_{W'})/\sqrt 3$.

The mass matrix $m_D$ is diagonal, and is parametrised by
$0<\eta<1$ such that
\begin{equation}
m_D =  M_{\em l} \eta S, \label {mddcpl}
\end{equation}
where
\begin{eqnarray}\label{S}
S &=& S(M_{W_R}) = {g_Rg_L\over8\pi^2}{1 \over 2\sqrt3 }
{M^2_{W_L} \over M^2_{W_R}}{m_b \over m_t} \ {\ln \left(
{M^2_{W_R} \over M^2_{W_L}} \right)} \nonumber \\ & \sim & 10^{-7}
\left( {\hbox{TeV} \over M_{W_R}}\right)^2.
\end{eqnarray}
\begin{figure}\label{fig1}
 \centering{
  \epsfig{file=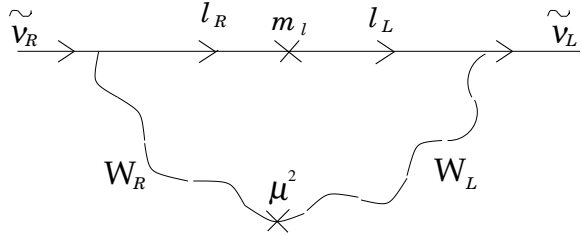,width=8cm}}
  \vskip0.4cm
\caption
  {Dirac mass generated by gauge interactions leading to the mass
term $\overline{\tilde \nu_L}\tilde \nu_R$. $\mu^2 \equiv g_R g_L
u_1 u_2$ is the $W_L - W_R$ mixing mass squared.}
\end{figure}
\noindent The mass mixing term $m_{\nu E}$ [Fig. \ref{fig2}]
arises at one-loop level via the gauge interactions

\begin{equation}
{g_L \over \sqrt 2} \overline {E^0_L} \not \! \! W^+_{L}E^-_L +
{g_R \over \sqrt 2} \overline {(E^-_R)^c}V^\dagger \not \! \!
W^+_{R}{\tilde \nu}_L + \hbox{H.c.}. \label{des}
\end{equation}
In contrast to the mass matrix $m_D$, the involvement of the
matrix $V$ in the interactions Eq. (\ref{des}) may mediate cross
generational mixing so that the matrix $m_{\nu E}$ is non-diagonal
in general. However, in the special case of decoupled generations
(i.e. $V \rightarrow {\bf I}$), the mass matrix $m_{\nu E}$
reduces to the diagonal form
\begin{eqnarray}
\lim_{V \rightarrow {\bf I}} m_{\nu E} = \eta M_E S. \label{mdnu}
\end{eqnarray}
\begin{figure}
  \centering
  \epsfig{file=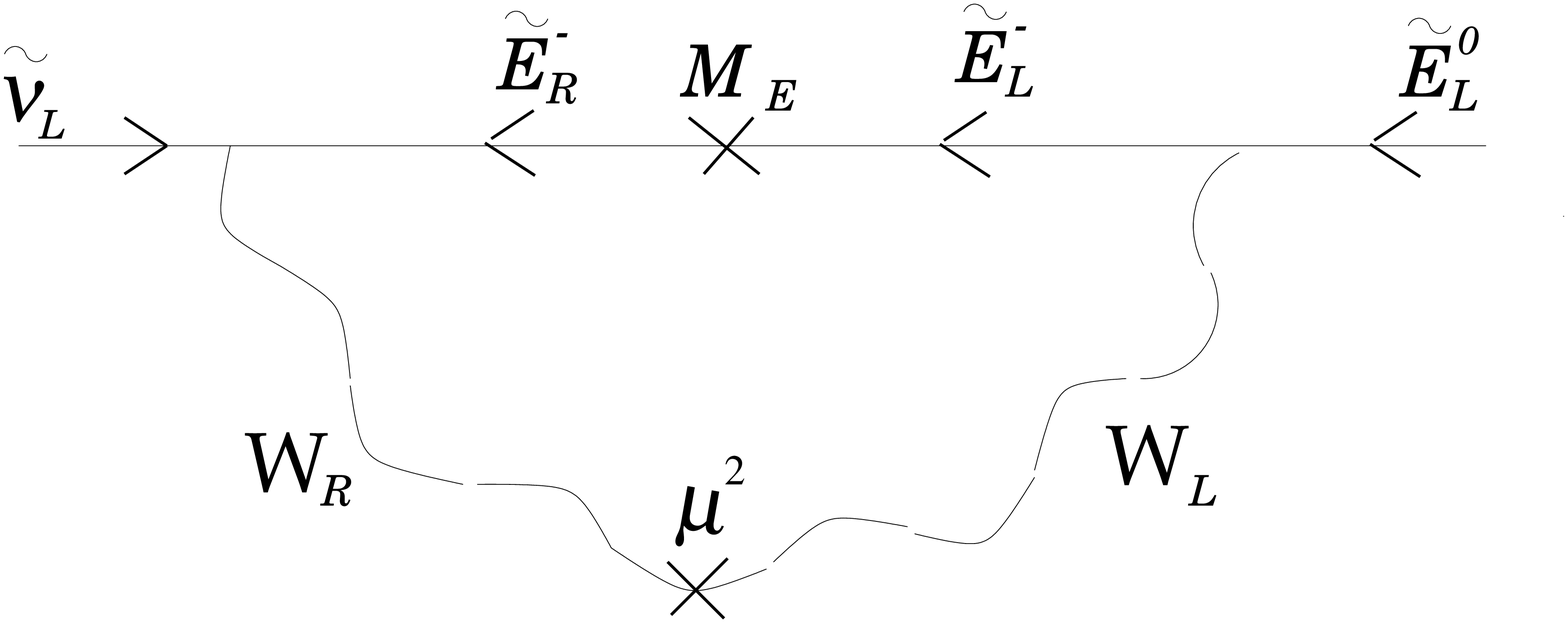,width=8cm}
  \vskip0.4cm
  \caption{$\overline{\tilde \nu_L}(\tilde E^0_L)^c$
neutrino mixing term generated by gauge interactions leading to
the mass term $m_{\nu E}$.}\label{fig2}
\end{figure}
\noindent Hence, we obtain an effective Lagrangian density for the
mass matrix of the neutrinos (which are approximately decoupled
from the heavy $E$ leptons),
\begin{eqnarray}
{\cal L}_{eff} &=& {1 \over 2} \left( \begin{array}{cccc}
\overline {{\tilde \nu}_L} & \overline {({\tilde \nu}_R)^c}
\end{array} \right) M_\nu \left( \begin{array}{cccc} ({\tilde
\nu}_L)^c \\ {\tilde \nu}_R \end{array} \right) + \hbox{ H. c.},
\label{Leff} \end{eqnarray} where the matrix $M_\nu$ is given by
\begin{eqnarray}
M_{\nu} \simeq \left(\begin{array}{cccc} 0 & m_D'
\\(m_D')^\dagger&M_R
\end{array}\right),
\label{Mnu}
\end{eqnarray}
with
\begin{equation}
m_D' = m_D + m_{\nu E} M_{E}^{-1} V^\dagger M_{\em l},
\label{mdprime}
\end{equation}
and $M_R$ is given in Eq. (\ref{MR}).

In the see-saw limit where the eigenvalues of $M_R$ are much
larger than the eigenvalues of $m'_D$ (which is generally valid),
the right and left neutrino states are effectively decoupled:
\begin{equation}
{\cal L}^{see-saw} \simeq {1 \over 2} \overline { \tilde{\nu}_L}
m_L ({\tilde \nu}_L)^c + {1 \over 2} \overline {({\tilde
\nu}_R)^c} M_{R} {\tilde \nu}_R + \hbox{ H.c.}, \label{Lseesaw}
\end{equation}
where
\begin{eqnarray}
m_L  &\simeq& - m_D' M_{R}^{-1}(m_D')^{\dagger}. \label{ml}
\end{eqnarray}
Knowledge of the mass matrix for the light neutrinos, $m_L$,
allows us to work out the oscillation parameters (i.e. the mixing
angles and $\delta m^2$) among the light neutrinos $\nu_L$, and
thereby make contact with the neutrino data. For the sake of
simplicity, we will drop the tilde in the neutrino fields in the
subsequent discussion whenever no confusion could arise. Unless
otherwise stated, the symbols $\nu_{L},\nu_{R}$ are all
(approximately) flavour eigenstates.

One way to obtain large $\nu_{eL}\rightarrow \nu_{\tau L}$
oscillations as suggested by the SNO data and other solar neutrino
experiments and small mixing angle ${\nu_{\mu L}} \rightarrow
{\nu_{eL}}$ oscillations for the LSND data is for the mass matrix
$m_L$ in Eq. (\ref{ml}) to have the approximate form
\begin{equation}\label{mlsolar}
    m_L \approx \pmatrix{0 & 0 &m_{1} \cr 0 & m_{2}& 0 \cr m_{1} & 0
    &0},
\end{equation}
where the `0' elements are not strictly zero, but much smaller
than the $m_i$. Clearly the mass matrix $m_L$ has 3 eigenvalues,
$\lambda'_1 \approx \lambda'_3 \approx m_1$, $\lambda'_2 \approx
m_2$. If this mass matrix is to be consistent with the solar and
LSND neutrino data, then it is required that:
\begin{description}
\item[A.]
$\lambda'_1$ and $\lambda'_3$ have to be split in such a way that
$|{\lambda'}_1^2 - {\lambda'}_3^2|$ is identified with $\delta
m^2_{solar}$,
\begin{equation}\label{ac1}
|{\lambda'}_1^2 - {\lambda'}_3^2| \equiv \delta m^2_{solar}.
\end{equation}
\item [B.]
For the sake of naturalness, the absolute mass scale of $m_{1}^2$
has to be much larger than $\delta m_{solar}^2$,
\begin{equation}\label{ac2}
m_{1}^2 \gg \delta m_{solar}^2.
\end{equation}
\item[C.]
 We require that
\begin{equation}\label{ac3}
\delta m^2_{LSND} = |{\lambda'}_1^2 - {\lambda'}_2^2| \simeq
|{m_1^2}-{m_{2}^2}|,
\end{equation}
where 0.2 eV$^2 \stackrel{<}{\sim}\delta m^2_{LSND}
\stackrel{<}{\sim} 10$ eV$^2$.
\end{description}

Our strategy now is to find the forms of the CKM matrices (and the
corresponding Yukawa matrices $\lambda_1 - \lambda_4$) which lead
to a neutrino mass matrix $m_L$ of the form Eq. (\ref{mlsolar}),
and also satisfy the requirements {\bf A, B} and {\bf C}. It turns
out that the above conditions do not lead to a unique solution.
However, if we impose an additional condition on the CKM-type
unitary matrices that:
\begin{description}
\item[D.]
There is a left-right similarity between the CKM matrix $K_L$ and
the corresponding CKM-type matrix $K_R$ for the SU(2)$_R$
interactions of the right-handed quarks, i.e.
\begin{equation}
K_L \approx K_R(\equiv Y_R^\dagger {K'}^{\dagger}) \approx {\bf
I}, \label{ac4}
\end{equation}
\end{description}
then apparently a unique picture emerges. That is, it is found
that there is a simple set of CKM-type matrices where the theory
is consistent with a $\hbox{SU(4)}\otimes\hbox{SU(2)}_L \otimes
\hbox{SU(2)}_R$ symmetry breaking scale of less than a few TeV and
reproduces the form of $m_L$ as in Eq. (\ref{mlsolar}) with the
conditions Eqs. (\ref{ac1})-(\ref{ac4}) satisfied:\footnote{The
form of these matrices could be derived following a similar
procedure as was done in Ref. \cite{422.3}. Recall that in Ref.
\cite{422.3} it was assumed that the atmospheric neutrino anomaly
is solved by $\nu_{\mu L}\rightarrow \nu_{\tau L}$ oscillations,
the LSND data is solved by ${\nu_{\mu L}} \rightarrow {\nu_{e L}}$
oscillations with the solar neutrino anomaly solved by $\nu_{e
L}\rightarrow \nu_{sterile}$ oscillations. While in the present
(post SNO) paper the solar neutrino anomaly is solved by $\nu_{e
L}\rightarrow \nu_{\tau L}$ oscillations, the LSND data is solved
by ${\nu_{\mu L} } \rightarrow {\nu_{e L}}$ oscillations and the
atmospheric neutrino anomaly is solved by $\nu_{\mu L}\rightarrow
\nu_{sterile}$ oscillations. The difference is essentially a
transformation of $\nu_{e L} \leftrightarrow {\nu_{\mu L}}$.}
\begin{eqnarray}\label{sch1m}
V \approx {\bf I_3}, \,\, U \approx \pmatrix{0 & 0 & 1 \cr 1 & 0 &
0 \cr 0& 1 & 0}, \,\, K' \approx Y_R'^\dagger \approx \pmatrix{1 &
0 & 0 \cr 0 & 0 & 1 \cr 0& 1 & 0}.
\end{eqnarray}
Eq. (\ref{ac4}) amounts to suggesting that the non-diagonal $K'$
(required for low symmetry breaking scale, see \cite{422.2}) and
non-diagonal $Y_R^\dagger$ (required to obtain large mixing angle
$\nu_{e L} \to \nu_{\tau L}$ oscillations) may have a common
origin.

The set of CKM-type matrices [Eq. (\ref{sch1m})] would arise in
the theory if we make the following ansatz for the Yukawa matrices
$\lambda_1, \lambda_3, \lambda_4$:
\begin{eqnarray}
\lambda_1 \approx \pmatrix{0&0&\times \cr \times&0&0 \cr
0&\times&0}, \ \lambda_3 \approx \pmatrix{\times &0&0 \cr
0&0&\times \cr 0&\times&0}, \ \lambda_4 \approx
\pmatrix{\times&0&0 \cr 0&0&\times \cr 0&\times&0},
\end{eqnarray}
with the matrix $\lambda_{2}$ approximately diagonal. In other
words, if $\lambda_1,\ldots,\lambda_4$ have the above form then a
low symmetry breaking scale ($\stackrel{<}{\sim}$ few TeV) is
phenomenologically viable, and additionally, the model can
accommodate the large mixing angle $\nu_{e L} \to \nu_{\tau L}$
solution to the solar neutrino problem as well as small angle
$\nu_{\mu L} \to \nu_{e L}$ oscillations as suggested by the LSND
experiment. In this solution scheme, the absolute scales of the
mass squared of the left-handed neutrinos turns out to
be\footnote{For numerical definiteness, we have assumed $S$ to
take on a value near its allowed upper bound, $S \sim 10^{-6}$,
see Eq. (\ref{S}).}
\begin{eqnarray}\label{massscale}
  m_{1}^2 \approx 16\eta^4 S^4
  M_{E_3}^2\bigg({m_e \over m_u}\bigg)^2  \sim 10^{-1} \eta^4
  \bigg({M_{E_3}\over \hbox{TeV}}\bigg)^2 \ \hbox{eV}^2, \nonumber \\
  m_2^2 \approx 4\eta^4 S^4
  M_{E_2}^2\bigg({m_\mu \over m_t}\bigg)^2 \sim
  10^{-6} \eta^4
  \bigg({M_{E_2}\over \hbox{TeV}}\bigg)^2 \ \hbox{eV}^2.
\end{eqnarray}

The matrices in Eqs. (\ref{ac4}),(\ref{sch1m}) are translated [via
the matrix $M_R$ in Eq. (\ref{MR})] into the following mass
spectrum for the right-handed neutrinos:
\begin{eqnarray}\label{sch1}
m_{\nu R1}\approx m_{\nu R3} &=& {m_\tau m_u \over M_{E_3}} + {m_e
m_c \over M_{E_1}} \sim {m_\tau m_u \over M_{E_3}}, \nonumber
\\ m_{\nu R2} &=& {2m_{\mu}m_{t} \over M_{E_2}}.
\end{eqnarray}
Among $M_{E_i}$, $M_{E3}$ is constrained by the requirement to
accommodate the solution to the LSND result [Eq. (\ref{ac3}) and
Eq. (\ref{massscale})] which implies the lower bound
\begin{equation}\label{Me3}
M_{E_3} \stackrel{>}{\sim} \hbox{TeV}.
\end{equation}
This bound on the mass of $M_{E_3}$ in turn constrains the mass
spectrum of $m_{\nu_R}$ to the ranges:
\begin{eqnarray}
1 \hbox{ keV} &\stackrel{<}{\sim} & m_{\nu R1} \approx m_{\nu R3}
\stackrel{<}{\sim} 10\hbox{ keV}, \nonumber \\  4 \hbox{ MeV}
&\stackrel{<}{\sim}& m_{\nu R2} \stackrel{<}{\sim} 1 \hbox{ GeV}.
\label{range}
\end{eqnarray}
In other words, in the present solution scheme there are two
approximately degenerate light right-handed neutrinos $\nu_{R1},
\nu_{R3}$ (with masses in the range keV $-$ 10 keV) and a heavy
$\nu_{R2}$ with mass in the range of $\sim$ MeV to $\sim 1$ GeV.

The mixing between the left and right neutrino sectors, though
suppressed, could be detected in a tell-tale kink (slope
discontinuity) in the otherwise continuous spectrum of a low
$Q$-value nuclear beta decay (such as the tritium beta decay)
\cite{kink}. For ordinary-sterile neutrino two-state mixing, the
weak eigenstates $\{\tilde \nu_{L},(\tilde \nu_{R})^c\}$ will be
linear combinations of the two mass eigenstates $\{\nu_{L},
(\nu_{R})^c\}$:
\begin{equation}\label{nuLmix}
\tilde \nu_{L} = \cos \psi \nu_{L} + \sin \psi(\nu_{R})^c, \,\,\,
(\tilde\nu_{R})^c = -\sin \psi \nu_{L} + \cos \psi(\nu_{R})^c,
\end{equation}
where the mixing angle between the $i$-left-handed state and
$j$-right-handed state is given by
\begin{equation}\label{sinLR}
  \sin^2 \psi_{ij} \sim \bigg| {m_i \over m_{\nu_{Rj}}} \bigg|,
\end{equation}
which is of significant interest only for $i j = \{13\},\{31\}$,
i.e.,
\begin{equation}\label{sinpsi}
\sin^2 \psi_{13} \simeq \sin^2 \psi_{31} \sim 10^{-3}\eta ^2
({\hbox{keV} / m_{\nu_{R1}}})^2.
\end{equation}
The mixing between $\tilde \nu_{\mu L}-(\tilde \nu_{R2})^c$ is
much smaller and we can ignore it (as with the other ${ij} \neq
\{13\},\{31\}$ channels). We will take the convention
\begin{equation}\label{sign}
\delta m^2_{ij} \equiv m^2_{\nu_{Rj}} - m^2_{i}
\end{equation}
so that $\delta m^2_{ij}$ is positive. Due to the mixing, the
spectrum of a weak decay that includes $\nu_{e L}$ in the final
state will consist of two components corresponding to $m_1$ and
$m_{\nu_{R3}}$. In the limit $m_{\nu_{R3}} \gg m_1$, the observed
beta spectrum can be expressed as the product of the massless
neutrino spectrum and a massive neutrino shape factor $S(E)$,
\begin{equation}\label{S1}
  {dN\over dE} \propto {dN(E,m_{1} = 0) \over dE}S(E),
\end{equation}
with
\begin{equation}\label{S2}
S(E) =   \left\{
            \begin{array}{cc}
            1 + \tan^2 \psi_{13} \bigg[ 1 - {m_{\nu_{R3}}^2 \over
       (Q-E)^2}\bigg]^{1/2} & \hbox{ for } E \leq Q-m_{\nu_{R3}} \\
            1               & \hbox{ for }E > Q - m_{\nu_{R3}},
          \end{array}
           \right.
\end{equation}
where $E$ is the beta decay's energy and $Q$ is the total decay
energy. The mixed spectrum will display a kink at $E = Q -
m_{\nu_{R3}}$ corresponding to the mass $m_{\nu_{R3}}$ and mixing
angle of Eq. (\ref{sinpsi}), which is a signature predicted by the
422 model. At present, the sensitivity of the experimental
searches for such kink in nuclear beta decays, which only sets an
upper bound of $\sin^2 \psi_{13} \stackrel{<}{\sim}$ few $10^{-3}$
for $m_{\nu_{R3}}$ in the range of a few keV \cite{PDBook}, is on
the verge of detecting the presence of the keV component predicted
by the 422 model. If the sensitivity of the experiments could
increase by two orders of magnitude in the future, we will be able
to verify (or falsify) the model by detecting (or not detecting)
this signature.

Having addressed the significance of the laboratory signature from
the mixing between the left- and right-handed neutrinos, we now
turn to investigate the cosmological implications of these
right-handed neutrinos which are potentially in conflict with the
standard cosmological energy density bound.\footnote{Note that the
exotic leptons $\{E^0, E^{-}\}$ do not lead to a cosmological
energy density problem. This is because their masses are heavy
enough to allow them to rapidly decay into quarks and leptons via
the gauge interactions.}

\section{The right-handed neutrinos in the framework of standard cosmology}

In this section we would like to determine if standard big bang
cosmology is consistent with the minimal 422 model. Within the
context of the standard big bang model, the present energy density
of a given particle species $X$, $\rho_X$ must not be much larger
than the critical density of the Universe $\rho_c$, see Eq.
(\ref{cosmic}). The first check of the implications of the minimal
alternative 422 model on the standard cosmological picture is
therefore to estimate the present energy density of the
right-handed neutrinos in this model. In the following we will
estimate the present day relic density of $\nu_{R1},\nu_{R3}$ with
the assumption that they are `hot' (i.e. particle species that
freeze-out from the thermal plasma while still relativistic) and
approximately stable in the standard early Universe scenario. The
heavier $\nu_{R2}$ neutrinos will be discussed separately in
Section IV.B.

In the alternative 422 model, the dominant decay mode of
$\nu_{R1}, \nu_{R3}$ is the $Z^0$-mediated tree-level process,
\begin{equation}\label{nrZll}
\nu_{R1,R3} \stackrel{Z^0}{\longrightarrow} \nu_{\tau, e} +
\bar{\nu}_{\alpha}\nu_\alpha \,\,\,(\alpha = e,\mu,\tau),
\end{equation}
with a lifetime of the order
\begin{equation}
\tau_{\nu_{R1,R3}} \sim {\tau_\mu \over \sin^2
\psi_{13}}\bigg({m_{\mu} \over m_{\nu_{R1}}}\bigg)^5 \sim {10^{22}
\over \eta^2}\bigg({\hbox{keV} \over m_{\nu_{R1}}}\bigg)^3
\hbox{s},
\end{equation}
(where $\tau_\mu \approx 10^{-6}$ s is the muon decay lifetime)
which is much larger that the age of the Universe, $t_U \sim
10^{17}$ s. Therefore $\nu_{R1},\nu_{R3}$ could be taken as
approximately stable as far as their contribution to the energy
density of the Universe is concerned.\footnote{In addition to the
tree-level decay of Eq. (\ref{nrZll}), there also exists a
sub-dominant radiative decay mode $\nu_{R}\rightarrow
\nu_{L}\gamma$ at one-loop level. Since the transition moment is
generated at one-loop level, its decay rate is suppressed. From
Ref. \cite{sanda} it can be estimated to be $\Gamma_\gamma
/\Gamma_\mu \sim {9\alpha \over 64 \pi}\sin^2 \theta_{mix}
(m_{\nu_R} / m_\mu)^5 (m_\tau / M_{W_L})^4\sin^2\phi \sim
10^{-43}\eta^2 \sin^2 \phi(m_{\nu_R} / \hbox{keV})^5$, where
$|\sin \theta_{mix}| = {\eta \over 2 \sqrt{3}}{M^2_{W_L} \over
M^2_{W_R}}{m_b \over m_t} \stackrel{<}{\sim} 10^{-4}$ is the
$W_L-W_R$ mixing angle and $\sin \phi$ is the relevant leptonic
$K_R$ mixing angle. The radiative lifetime of $\nu_{R}\rightarrow
\nu_{L}\gamma$ should be larger than $10^{24}$ s, which is
required by the astrophysical constraint from the diffuse photon
background for a radiatively decaying species $X$ in the mass
range of $m_X {\sim}$ keV \cite{Kolb}. Not surprisingly, we find
that this is in fact the case for the light right-handed neutrinos
for all parameter space of interest, and thus this decay mode can
be safely ignored.}

The relic density is a function of the masses of the right-handed
neutrinos. Obviously, a heavier relativistic particle has a larger
contribution to the relic density. Because various observations
limit the relic density to be less than the critical density of
the Universe, this will in turn suggest an upper limit on the
masses of these particles which can be compared with the mass
range of Eq. (\ref{range}). It is fairly straightforward to
estimate the constraint imposed on the masses of the right-handed
neutrinos from $\Omega_{\nu_{R}}h^2 \stackrel{<}{\sim} 1$ within
the standard cosmological framework (see for example Chapter 5 in
Ref.\cite{Kolb}).

In order to find out the relic abundance of the lightest
right-handed neutrino we need to estimate its production rate.
Production of right-handed neutrinos can occur via two distinct
mechanisms. First there is a direct production by $Z'$- and
$W_R$-mediated interactions, such as\footnote{In addition to the
process in Eq. (\ref{therma}), there are also other channels that
contribute to the production of the right-handed neutrinos,
including $W_R$ mediated processes. The contribution from $W_R$
mediated exchange will be at most the same order to that of the
$Z'$ exchange channel. For simplicity sake, in the following
calculation we shall only take the $Z'$ exchange channel into
account for estimating the production of $\nu_R$. Ignoring the
$W_R$ channel would not effect the conclusions of the present
paper.}
\begin{eqnarray}
e_R \overline{e_R} \stackrel{Z'}{\longleftrightarrow} \overline
{\nu_R} {\nu_R}. \label{therma}
\end{eqnarray}
Second, right-handed neutrinos can be produced via oscillations of
$\nu_L \leftrightarrow (\nu_R)^c$. However, for the purpose of
this section, we will ignore the second production mechanism
because it will not modify our general conclusions.

To estimate the relic abundance of the $\nu_{R1},\nu_{R3}$ from
direction production, we have to sum over the averaged direct
production rates, i.e. $\Gamma = \sum_\psi \langle \Gamma_{\bar
\psi \psi \rightarrow \overline{\nu_R} \nu_R}\rangle$ for $\psi =
e_L, e_R, \nu_{\alpha L}$. \vskip0.4cm
\begin{figure}
  \centering
  \epsfig{file=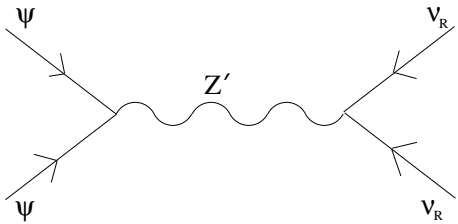,width=6cm}
  \vskip 0.4 cm
  \caption{Direct production of the right-handed neutrinos via
  scatterings of $\psi
\overline{\psi} {\leftrightarrow} \overline{\nu_R} {\nu_R}$
mediated by $Z'$, where $\psi = e_L, e_R,\nu_{\alpha L}$ ($\alpha
= e, \mu, \tau$).}\label{eerr}
\end{figure}
\noindent As the Universe expands, the temperature (denoted by
$T$) decreases, and so does the rate of direct right-handed
neutrino production. When the temperature drops below a certain
value $T_F$ (the freeze-out temperature) the production rate would
not be large enough to thermalise the right-handed neutrinos with
the thermal background. The right-handed neutrino is said to
freeze-out from the thermal background, leaving behind its
abundance frozen at the value it had when last in thermal
equilibrium. The freeze-out temperature is approximately
determined from the condition that the scattering rate $\Gamma$
has decreased to that of the Hubble expansion rate $H$,
\begin{equation}\label{tf0}
 \Gamma(T) \approx H(T),
\end{equation}
where the Hubble expansion rate is given by
\begin{equation}\label{Hubble}
H = 1.66 \sqrt{g_\ast}{T^2/M_{Pl}},
\end{equation}
and the scattering rate is given by
\begin{equation}\label{Gamma}
\Gamma(T) = \sum_{\psi}\langle \Gamma_{{\bar \psi} \psi
\rightarrow \overline{\nu_R} \nu_R }\rangle \approx
{1\over4}{G_F'}^2 T^5.
\end{equation}
In Eq. (\ref{Hubble}), $M_{Pl}$ is the Planck mass and $g_\ast$
counts the relativistic degrees of freedom that contribute to the
energy density of the Universe, as defined in the usual way, by
$\rho_R = {\pi \over 30}g_\ast T^4$. In Eq. (\ref{Gamma}),
${G_F'}$ is a Fermi constant-like quantity that characterises the
strength of the $Z'$ neutral current interaction,
\begin{equation}
 G'_{F} = \bigg({M_{W_L} \over M_{Z'}}\bigg)^2 G_F,
\end{equation}
where $G_F \simeq 10^{-5}$ GeV$^{-2}$ is the Fermi constant. The
freeze-out temperature $T_F$ can be estimated by solving Eq.
(\ref{tf0}), which gives
\begin{equation}\label{tf}
T_F \sim 50 \hbox{ MeV} \,\,\,\hbox{for } M_{Z'} \sim \hbox{TeV}.
\end{equation}
Comparing Eq. (\ref{tf}) with the mass spectrum Eq. (\ref{range})
we see that the light right-handed neutrinos are indeed
relativistic at $T \approx T_F$, which is what we have assumed.

\par The right-handed neutrinos, after
freezing out from the thermal background, would still maintain
their equilibrium distribution at temperature $T_{\nu_R}$ which
eventually becomes smaller than the background photon temperature
$T$ by a factor of
\begin{equation}\label{entransfer}
{T_{\nu_R}\over T} = \bigg({{g_{\ast S}^{after}} \over {g_{\ast
S}^{before}}}\bigg)^{1/3} \stackrel{<}{\sim} 1
\end{equation}
when the entropy from the $e^{\pm}$ annihilation is transferred to
the photons at $T\approx m_e$. In Eq. (\ref{entransfer}), $g_{\ast
S}^{before}$ counts the total number of relativistic degrees of
freedom that contribute to the entropy density $s$ of the Universe
(as defined by $s = {2 \pi^2 \over 45}g_{\ast S} T^3$) just a
moment before $T = m_e$, while $g_{\ast S}^{after}$ counts the
number of relativistic degrees of freedom at some later time.
Standard calculation for the present energy density of the
$\nu_{R1},\nu_{R3}$ right-handed neutrinos (normalised to the
critical density) leads to
\begin{eqnarray}
\Omega_{\nu_{R1}+\nu_{R3}} h^2 = {n_{\nu_{R}}m_{\nu_{R1}}h^2 \over
\rho_c} = {3 \over 4}\bigg({g_{\ast S}^{after} \over g_{\ast
S}^{before}}\bigg) {n_\gamma m_{\nu_{R1}} h^2 \over \rho_c},
\end{eqnarray}
where ${n_{\nu_{R}} \over n_\gamma} = {3\over 4}({g_{\ast
S}^{after} \over g_{\ast S}^{before}})$ is the ratio of the number
density of the $\nu_R$'s to that of the photons in the cosmic
microwave background (CMBR). At present, $n_\gamma = 422$
cm$^{-3}$. The constraint $\Omega_{\nu_{R1}+\nu_{R3}} h^2
\stackrel{<}{\sim} 1$ restricts $m_{\nu_{R1}}$ to the range
\begin{equation}\label{mres}
  m_{\nu_{R1}} \stackrel{<}{\sim} 30 \ {g_{\ast S}^{before} \over g_{\ast
S}^{after}} \hbox{ eV},
\end{equation}
which is clearly not consistent with the mass spectrum of the
right-handed neutrinos in Eq. (\ref{range}), as the factor
$({g_{\ast S}^{before}}/{g_{\ast S}^{after}})$ is constrained to
be around $16/5.82 = 2.75$ for $M_{Z'} {\sim}$ few TeV. Therefore
we are led to the conclusion that, in the framework of standard
cosmology, even the lightest right-handed neutrinos are not
consistent with the cosmological energy density bound of Eq.
(\ref{cosmic}). They will over-close the Universe.

Confronted with the inconsistency with the cosmological energy
density bound, one could attempt to modify the particle physics to
get around the conflict with standard cosmology. A popular way to
do this is by introducing a massless Majoron $J$ that arises from
the spontaneous breaking of an imposed global symmetry. Things
could be arranged in such a way that the coupling of $J$ with
$\nu_R$ opens up an invisible decay channel of $\nu_R$ into the
undetected Majoron (and $\nu_L$) that is rapid enough to alleviate
the cosmological bounds. However, in our opinion, such a remedy is
rather desperate, and we would not pursue it further to avoid
spoiling the elegance of the 422 model. Instead of modifying the
model we prefer to explore an alternative cosmology scenario to
find a way out of the conflict. This will be done in the next
section.

\section{Right-handed neutrinos in the framework of
non-standard cosmology with low reheating temperature}

It is usually assumed that the radiation-dominated era commences
after a period of inflation, and that the cold Universe at the end
of inflation becomes the hot Universe of the radiation-dominated
era in a process known as reheating. During reheating, a thermal
bath of relativistic particles (e.g. electrons and photons) is
slowly formed as the coherent oscillations of a condensate of
zero-momentum massive scalar field $\phi$ decays
\cite{Kolb}. The completion of the $\phi$ decay
marks the commencement of radiation-dominated era at an initiation
temperature $T_{RH}$. From a phenomenological point of view, the
reheating temperature $T_{RH}$, which is given in terms of
$\Gamma_\phi$, the lifetime of the massive scalar field $\phi$
\cite{Kolb},
\begin{equation}\label{Tauphoi}
    T_{RH} = \sqrt{M_{Pl} \Gamma_\phi}\left[{90 \over 8 \pi^3
  g_\ast(T_{RH})}\right]^{1/4},
\end{equation}
could be treated as a free parameter that is model-dependent (i.e.
via the dynamics of the $\phi$ physics in an expanding cosmic
background). It is an {\em a priori} assumption in standard big
bang cosmology that at the initiation of the radiation-dominated
era, thermal and chemical equilibrium prevail as an initial
condition, which is equivalent to the hypothesis that $T_{RH}$ is
higher than the freeze-out temperature of the cosmological process
under consideration [in our case here the pertinent process is the
production of the right-handed neutrinos, with the freeze-out
temperature $T_F \sim 50 \hbox{ MeV}$]. However, there is no
empirical evidence of the radiation-dominated era before the epoch
of BBN, i.e. temperature above $\sim$ 1 MeV. The only real
constraint on $T_{RH}$ is that suggested by BBN which implies that
$T_{RH}$ could be as low as 0.7 MeV
\cite{Kawasaki:1999na,Kawasaki:2000en}.

If the reheating temperature is indeed only of order $\sim$ MeV,
interesting modifications to some standard cosmological bounds on
particle physics would be necessitated. For example, with such a
low reheating temperature scenario a given dark matter species $X$
may never achieve chemical equilibrium with the thermal radiation
background, resulting in a relic abundance that is much lower than
that predicted by standard cosmology. It may therefore be possible
that the 422 model might be reconciled with cosmology if $T_{RH}$
is low enough (the calculation of the previous section only holds
in the limit of high $T_{RH}$). We now study this possibility.

\subsection{Relic abundance of the
light $\nu_R$ in the low
reheating Universe}

In this subsection we would like to answer the question of whether
the alternative 422 model is consistent with the low reheating
Universe by first estimating the relic density of the $\nu_{R1},
\nu_{R3}$ neutrinos produced via collisional processes in the
non-standard cosmological scenario. In such a low reheating
scenario,
\begin{equation}\label{trhlarge} T_{RH}
< T_F \sim 50 \hbox{ MeV},
\end{equation}
which implies that chemical equilibrium of $\nu_R$ is not attained.
This means that  $n_{\nu_R} \ll n_{\nu_R}^{eq}$ (where $n_{\nu_R}$,
$n^{eq}_{\nu_R}$ are the actual and equilibrium densities). The
heavier $\nu_{R2}$ right-handed neutrino will be treated
separately in Section IV.B.

There are two distinct types of collisional processes. First, we
have the direct production, as already considered in Section III.
Second, we have the effect of collisions on the oscillating
neutrino ensemble. Our purpose is to obtain the present day
abundance of the right-handed neutrinos by integrating the
corresponding Boltzmann equation that governs the time evolution
of these production mechanisms.

Assuming $n_{\nu_R} = n_{\overline{\nu_R}}$, the Boltzmann
equation pertinent to the production of the right-handed neutrinos
via collisional processes in an expanding cosmic background is
given by \cite{Kolb}
\begin{equation}\label{bz1}
{dn_{\nu_R} \over dt} + 3 H n_{\nu_R} = - \langle \sigma| v|
\rangle \big[ n_{\nu_R}^2 - (n_{\nu_R}^{eq})^2 \big],
\end{equation}
where $\langle \sigma |v| \rangle$ is the total thermal averaged
cross section times velocity for the relevant collisional process
that produces the $\nu_R$ state. It is useful to scale out the
effect of the expansion of the Universe by considering the
evolution of the number of particles in a comoving volume
$Y_{\nu_R} \equiv n_{\nu_R} /s$, where $s$ is the entropy density.
Introducing an independent parameter that explicitly depends on
the temperature $T$, $x \equiv m_{\nu_R}/T$, Eq. (\ref{bz1}) can
be expressed in the form
\begin{eqnarray}\label{bz2}
    {x \over Y_{\nu_R}^{eq}}{dY_{\nu_R} \over dx} &=& - {\Gamma \over H}
   \bigg[  \bigg( {Y_{\nu_R} \over Y_{\nu_R}^{eq}}\bigg) ^2 - 1 \bigg],
\end{eqnarray}
where $\Gamma \equiv n_{\nu_R}^{eq}\langle \sigma |v|\rangle$. If
the reheating temperature is lower than $T_F \sim 50$ MeV (or
equivalently, $\Gamma \ll H $), then the interactions are too weak
for chemical equilibrium to be attained. In this case, the
$Y_{\nu_R}$ term in the right-handed side of Eq. (\ref{bz2}) could
be dropped (to a good approximation) and the equation recast
into the form
\begin{eqnarray}\label{y}
{4x \over 3 Y_{\gamma}} {dY_{\nu_R} \over dx} \simeq {\Gamma \over
H}.
\end{eqnarray}
In arriving at Eq. (\ref{y}) we have made use of the ratio of
fermion ($\nu_R$) and boson (photon) at thermal equilibrium,
$Y^{eq}_{\nu_R}/Y_\gamma = n^{eq}_{\nu R} / n_\gamma = {3 / 4}$,
where the possible factor of $(T_{\nu R} / T)^3 =
g_{after}^\ast/g_{before}^\ast$ (which is not significantly
different than 1) is ignored.

In principle, to evaluate the present density of the right-handed
neutrinos we have to integrate Eq. (\ref{y}) for both reheating
era plus radiation-dominated era, i.e. $\int dY_{\nu R} =
\int_{reheating} dY_{\nu R} + \int_{radiation} dY_{\nu R}$.
However, $\int_{reheating} dY_{\nu R}$ is governed by
model-dependent physics of reheating and its contribution to the
production of the right-handed neutrinos is not determined with
definiteness. We therefore consider only the right-handed
neutrinos produced in the radiation-dominated era (i.e. for $T
\leq T_{RH}$) and approximate $\int dY_{\nu R} \approx
\int_{T_{RH}} ^{T_0} dY_{\nu R}$, in which the Hubble expansion
rate $H$ scales with temperature as in Eq. (\ref{Hubble}).

Hence the present day relic abundance of the $\nu_{R1}, \nu_{R3}$
neutrinos is taken as $Y_{\nu_{R0}} \approx Y_{\nu_R}(T =
m_{\nu_{R1}})$, obtained by integrating Eq. (\ref{y}) from $T =
T_{RH}$ to $T=m_{\nu_{R1}}$ by employing the appropriate
relativistic form of the `effectiveness of production' term,
$\Gamma / H$\footnote{The contribution to $Y_{\nu_R}$ from the
non-relativistic regime $T = m_{\nu_{R1}} \to T = T_0$ is very
tiny compared to that from the relativistic regime, and hence can
be ignored.}. The collisional production rate, $\Gamma \equiv
\Gamma^{col}$ is the sum of the direct production term and a
decoherence term (which arises for the effects of collisions on
the oscillating neutrino ensemble),
\begin{equation}\label{GG}
  \Gamma^{col} = \Gamma^{dp} + \Gamma^{col-os}.
\end{equation}
As was considered in Section III, the relativistic form of
$\Gamma^{dp}$ for the direct production mechanism is given by Eq.
(\ref{Gamma}). The production rate, $\Gamma^{col-osc}$, is given
by \cite{Langacker:1989sv}
\begin{eqnarray}\label{Gcoll}
  \Gamma^{col-os} &=&  (D_{\nu_{e}} + D_{\nu_\tau}) \langle
  \sin^2 ({\tau_{coll}
/ \tau_{osc}}) \rangle \sin^2 2\psi \nonumber \\ &\simeq& {y
G^2_FT^5\over 2}{1 \over 2} \sin^2 2\psi,
\end{eqnarray}
where \cite{HStodolsky}
\begin{eqnarray}\label{D}
D_{\nu_\alpha} = {1 \over 2}\Gamma_{\nu_{\alpha}} \simeq {1\over
2} y_{\nu_\alpha} G^2_FT^5
\end{eqnarray}
is the thermally averaged collision frequencies between the
left-handed $\alpha$-neutrinos with the background plasma, and $y
= y_{\nu_e} + y_{\nu_\tau} = 4.0 + 2.9 = 6.9$ \cite{enqvist}. The
time scales characterising the period of oscillation of
$\nu_{L}-(\nu_{R})^c$, $\tau_{osc}$, and that of the collisions
between the $\nu_L$'s with the thermal background, $\tau_{coll}$,
are given by
\begin{eqnarray}
\langle \tau_{osc} \rangle &\equiv& {4 \langle E \rangle \over
\delta m_{ij}^2} \sim 10^{-14} \bigg({\hbox{keV} \over
m_{\nu_{R1}}}\bigg)^2 \bigg({T \over \hbox{MeV}}\bigg) \hbox{ s},
\nonumber \\ \langle \tau_{coll} \rangle &\sim&
 1 / D_{\nu_\alpha} \sim \hbox{few} \bigg({\hbox{ MeV}
\over T}\bigg)^5\hbox{ s}.
\end{eqnarray}
Clearly, the oscillation period is very short compared to the
average interval between collisions, i.e. $\langle \tau_{osc}
\rangle \ll \langle \tau_{coll} \rangle$ for all of the
temperature range of interest, so that many oscillations occur
between successive collisions. As a result, the collisions will
not significantly destroy the coherence of the ensemble and we
have $\langle \sin^2 ({\tau_{coll} / \tau_{osc}}) \rangle = {1
\over 2}$ in Eq. (\ref{Gcoll}).

Because the right-handed neutrinos are much heavier than the
left-handed states (so that the sign of $\delta {m_{ij}^2} \equiv
m^2_{\nu_{Rj}} - m^2_{i} \approx + m^2_{\nu_{Rj}}$ as defined via
Eq. (\ref{sign}) is positive), no oscillation neutrino asymmetry
amplification can occur via the oscillations
\cite{ftv96,fv1prdyr97}. Furthermore, since $\delta m^2_{ij}
\stackrel{>}{\sim} 10^6$ eV$^2$ [given the mass range of
$m_{\nu_R}$, Eq. (\ref{range})], it turns out that we can neglect
matter effects for the temperature range of interest\footnote{In
general, due to interactions that discriminate between the active
and sterile neutrinos by the thermal background (i.e. the matter
effect) \cite{Wolfenstein:1978ue,Barger80}, an effective potential
\cite{Notzold} will be induced and felt by the $\alpha$-flavour
active neutrinos, thus changing the dynamics of the coherent
oscillations. The effective potential can be put into the
convenient form \cite{fv1prdyr97}
\begin{eqnarray}
V_\alpha = (- a + b){\delta m^2_{ij} \over 2(3.15 T)},
\nonumber
\end{eqnarray}
where the dimensionless variables $a$ and $b$ are given by:
\begin{eqnarray}
  a = -2.5 \times 10^{-5}L^{(\alpha)}\bigg({\hbox{keV}^2 \over
  \delta m^2_{ij}}\bigg)    \bigg({T \over \hbox{MeV}}\bigg)^4,
  \, \, b = - 10^{-6} \bigg({T \over 13 \hbox{ MeV}}\bigg)^6
         \bigg({\hbox{keV}^2 \over \delta m^2_{ij}}\bigg).
\nonumber
\end{eqnarray}
Here $L^{(\alpha)}$ is related to the lepton number asymmetry associated
with
the $\alpha$-flavour active neutrino. If $L^{(\alpha)}$ is not too
big then $|a|,|b| \ll 1$ for the temperature range of interest,
and hence the matter effect is very tiny and can be neglected.}.
This justifies the use of Eq. (\ref{Gcoll}).

We can now plug in the collision rates for both collisional
mechanisms, Eq. (\ref{Gamma}) and Eq. (\ref{Gcoll}) into Eq.
(\ref{y}) for $\Gamma$ and integrate it to obtain the present
number densities of the $\nu_{R1}, \nu_{R3}$ right-handed
neutrinos,
\begin{eqnarray}\label{yygamma}
   {n^{dp}_{\nu_{R0}} \over n_\gamma}
   \sim 10^{-6}\bigg({\hbox{TeV} \over M_{Z'}}\bigg)^4
  \bigg({T_{RH} \over \hbox{MeV}}\bigg)^3, \nonumber \\
{n^{col-osc}_{\nu_{R0}} \over n_\gamma}
   \sim 10^{-4} \eta^2\bigg({\hbox{keV} \over
   m_{\nu_{R1}}}\bigg)^2
  \bigg({T_{RH} \over \hbox{MeV}}\bigg)^3.
\end{eqnarray}
Clearly the number density of the right-handed neutrinos is very
low, which is of course expected given that $T_{RH} < T_F$. The
total contribution of the right-handed neutrinos produced via the
collisional mechanisms normalized to the critical energy density of the
Universe is easily computed to be
\begin{eqnarray}\label{omcol}
\Omega^{col}_{\nu_{R1}+\nu_{R3}}h^2 =
\Omega^{dp}_{\nu_{R1}+\nu_{R3}}h^2 +
\Omega^{col-osc}_{\nu_{R1}+\nu_{R3}}h^2,
\end{eqnarray} where
\begin{eqnarray}
\Omega^{dp}_{\nu_{R1}+\nu_{R3}}h^2 \sim 10^{-5}
\bigg({m_{\nu_{R1}}\over \hbox{keV}}\bigg) \bigg({\hbox{TeV} \over
M_{Z'}}\bigg)^4 \bigg({T_{RH} \over \hbox{MeV}}\bigg)^3
\end{eqnarray}
and
\begin{eqnarray}
\Omega_{\nu_{R1}+\nu_{R3}}^{col-osc}h^2 \sim 10^{-2}\eta^2
\bigg({T_{RH} \over \hbox{MeV}}\bigg)^3 \bigg({\hbox{keV} \over
m_{\nu_{R1}}}\bigg).
\end{eqnarray}
We see that $\Omega^{col}_{\nu_{R1}+\nu_{R3}}h^2$ is consistent
with a value of $\stackrel{<}{\sim}$ unity for a significant range
of parameters
\footnote{Technically, Eq. (\ref{omcol}) is not complete. In
addition to the collisional processes, there is the effect of
`oscillations between collisions'. Because $\langle \tau_{osc}
\rangle \ll \langle \tau_{coll} \rangle$, the number density of
$\nu_R$ from oscillation of $\nu_L$ is simply given by
\begin{eqnarray}
\nonumber
  {n^{osc}_{\nu_{R}}} \approx n_{\nu} \cdot {1\over 2} \sin^2 2\psi,
\end{eqnarray}
where $n_{\nu}$ is the number density of the relic $\nu_L$
neutrino background, which is related to $n_\gamma$ via $n_\nu /
n_\gamma = 3/11$ at present. Thus, the energy density of the light
right-handed neutrinos produced via the `pure oscillation' channel
is given by
\begin{eqnarray}
\nonumber
  \Omega^{osc}_{{\nu_{R1}+\nu_{R3}}}h^2 \sim {3 n_\gamma
  \over 11} {m_{\nu_{R1}} h^2 \over \rho_c}{\sin^2 2\psi \over 2}
  \sim {10^{-2}\eta^2}
  \bigg({\hbox{keV} \over m_{\nu_{R1}}}\bigg).
\end{eqnarray}
Clearly this pure oscillation contribution is within the
cosmological bound for all parameter space of interest.}

\begin{equation}\label{Tm}
  T_{RH} \stackrel{<}{\sim}50 \bigg[ \bigg({m_{\nu_{R1}}\over
  \hbox{keV}}\bigg) \bigg({\hbox{TeV}\over{M_{Z'}}}\bigg)^4
  + 10^3 \eta^2 \bigg({\hbox{keV}\over m_{\nu_{R1}}}\bigg)
  \bigg]^{-{1\over3}}\ {\hbox {MeV}}.
\end{equation}
We thus conclude that the relic abundance of the
$\nu_{R1},\nu_{R3}$ neutrinos is consistent with the cosmological
energy density bound in the low reheating scenario.

Let's summarise what we have done so far for the
${\nu_{R1},\nu_{R3}}$ neutrinos. In the alternative 422 model
developed to accommodate the LSND and solar neutrino anomalies,
the masses $m_{\nu_{R1}} \approx m_{\nu_{R3}}$ lie in the range
given by Eq. (\ref{range}), that is 1 keV $\stackrel{<}{\sim}
m_{\nu_{R1}},m_{\nu_{R3}} \stackrel{<}{\sim} 10$ keV. In the limit
of high reheating temperature, $T_{RH} \to \infty$, the
right-handed neutrinos are fully populated and, as shown in
Section III, would be inconsistent with the cosmological energy
density bound. However, in a low reheating scenario, $n_{\nu_R}$
is suppressed and these right-handed neutrino will be consistent
with the cosmological energy density bound provided that Eq.
(\ref{Tm}) holds.

Note that since their existence is not excluded by the
cosmological energy density bound in a low reheating Universe, the
light right-handed neutrinos ${\nu_{R1},\nu_{R3}}$, being
effectively stable, could be a viable dark matter candidate.
Specifically, the right-handed neutrinos in the range of keV could
play a role as warm dark matter \cite{WDM}. We will leave the
details of this possibility for future study.
\subsection{Heavy right-handed neutrino decay mediated via the $W_R$ gauge
boson}

Having shown that the two lightest right-handed neutrinos, $
\nu_{R1},\nu_{R3}$, could exist in a low reheating Universe
without violating the cosmological energy density bound, we are
still left with the heavy right-handed neutrino, $\nu_{R2}$, to
worry about. However, because $\nu_{R2}$ has a large mass (4 MeV
$\stackrel{<}{\sim} m_{\nu_{R2}} \stackrel{<}{\sim}$ 1 GeV), it
can decay much more rapidly than $\nu_{R1},\nu_{R3}$. Furthermore
its production rate is highly suppressed if $T_{RH} \ll
m_{\nu_{R2}}$. For simplicity, we first consider $\nu_{R2}$ in the
standard case of high reheating temperature. Of course, we should
keep in mind that in the case of low reheating temperature the
constraints will be much weaker.

\begin{figure}
  \centering
  \epsfig{file=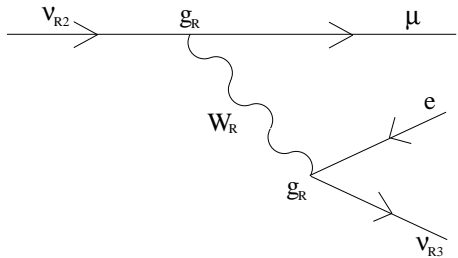,width=6cm}
  \vskip0.4cm
  \caption{$\nu_{R2} {\rightarrow}
\mu^- e^+ \nu_{R3}$ decay channel dominated by
$W_R$.}\label{nrdecay}
\end{figure}

If $\nu_{R2}$ decays rapidly enough, then it will not lead to any
cosmological problems. If kinematically allowed (i.e.
$m_{\nu_{R2}} > m_{\mu} + m_e + m_{\nu_{R3}}$), the dominant decay
channel of $\nu_{R2}$ is
\begin{equation}\label{nrdcaymu}
\nu_{R2} \stackrel{W_{R}}{\longrightarrow} \mu^- {e^+} \nu_{R3}.
\end{equation}
This Feynman diagram is shown in Fig. \ref{nrdecay}. \vskip0.4cm
\noindent The decay of $\nu_{R2} \stackrel{W_R}{\longrightarrow}
\mu^- e^+ \nu_{R3}$ is similar to that of $\mu^-$ decaying into
$e^{-} \nu_{\mu} + \bar \nu_{e}$, with $\nu_{R2}$ playing the role of
$\mu^-$.
In the limit where the masses of the decay products vanish in
comparison to $m_{\nu_{R2}}$, the lifetime of the $\nu_{R2}
\stackrel{W_R}{\longrightarrow} \mu^- e^+ \nu_{R3}$ decay can be
roughly expressed in terms of the muon lifetime, $\tau_\mu \approx
10^{-6}$ s,
\begin{eqnarray}
&& \tau_{\nu_{R2}} \approx \tau_{\mu}\bigg( {g_L \over
g_{R}}\bigg)^4 \bigg( {M_{W_R} \over M_{W_L} }\bigg)^4 \bigg(
{m_{\mu} \over m_{\nu_{R2}} }\bigg)^5 = 0.2 \bigg({M_{W_R} \over
\hbox{TeV}}\bigg)^4 \bigg( {m_{\mu} \over m_{\nu_{R2}} }\bigg)^5
\hbox{ s} \nonumber
\\ && \hbox{for } m_{\nu_ {R2}} \stackrel{>}{\sim} m_{\mu}+ m_e +
m_{\nu_{R3}},
\label{tauR2}
\end{eqnarray}
which is short enough to be consistent with all cosmological and
astrophysical bounds.

If $m_{\nu_{R2}} \stackrel{<}{\sim} m_\mu$ then the decay of
$\nu_{R2}$ into muon is not allowed. In this case $\nu_{R2}$
decays via cross generational mixing into $e^- e^+ \nu_{R3}$ which
is suppressed by a cross generational mixing angle $|\sin
\phi_{12}| \stackrel{<}{\sim} 10^{-1}$. Explicitly, the lifetime
of $\nu_{R2} \to e^- e^+\nu_{R3}$ is
\begin{eqnarray}\label{nu2ee}
\tau_{\nu_{R2}\to e^- e^+\nu_{R3}} &=& \tau_{\mu}\bigg( {g_L^2
\over g_{R}^2\sin\phi_{12}}\bigg)^2 \bigg( {M_{W_R} \over M_{W_L}
}\bigg)^4 \bigg( {m_{\mu} \over m_{\nu_{R2}} }\bigg)^5 \nonumber
\\ &\approx &3 \times 10^{11} \bigg({10^{-1} \over \sin
\phi_{12}}\bigg)^2 \bigg({M_{W_R} \over \hbox{TeV}}\bigg)^4
\bigg({\hbox{MeV} \over m_{\nu_{R2}} }\bigg)^5 \hbox{ s}.
\end{eqnarray}
Because the annihilation of the $e^+e^{-}$ pair from this decay
mode could produce photons that potentially distort the CMBR, the
$\nu_{R2} \to e^- e^+ \nu_{R3}$ decay channel is subjected to the
stringent CMBR constraint $\tau_{\nu_{R2}\to e^- e^+\nu_{R3}}
\stackrel{<}{\sim} 10^6$ seconds [see Fig. 5.6 in Ref.
\cite{Kolb}]. Using Eq. (\ref{nu2ee}), this constraint implies
that
\begin{equation}\label{phi12}
\sin \phi_{12} \stackrel{>}{\sim} 10^{-1} \bigg({10 \hbox{MeV}
\over m_{\nu_{R2}}}\bigg)^{5/2}\bigg({M_{W_R}\over
\hbox{TeV}}\bigg)^2.
\end{equation}
In summary, for $m_{\nu_{R2}} \stackrel{>}{\sim} m_\mu$ the decay
of $\nu_{R2}$ is rapid enough to be consistent with standard
cosmology (irrespective of the value of the reheating
temperature). For $m_{\nu_{R2}} \stackrel{<}{\sim} m_\mu$ the
decay is rapid enough provided that Eq. (\ref{phi12}) holds.
Recall that this is only valid in the limit where $T_{RH}$ is
high. In the case of low $T_{RH}$, the astrophysical constraint is
much weaker (depending on $T_{RH}$).

\section{Conclusion}
The alternative 422 model is an interesting extension to the
standard model for many reasons, which include addressing the
problem of neutrino masses and their mixings. In its minimal form
the model quite naturally accommodates a set of simultaneous
solutions (with active-active neutrino oscillations) to the LSND
and solar neutrino data without involving any mass scale higher
than a few TeV - thereby avoiding the hierarchy problem. It turns
out that the masses of the 3 right-handed neutrinos $\nu_{Rj}$ $(j
= 1,2,3)$ are constrained to lie between 1 keV $-$ 10 keV (for
$\nu_{R1}, \nu_{R3}$) and 4 MeV $-$ 1 GeV (for $\nu_{R2}$) [see
Eq. (\ref{range})].

On the other hand, standard hot big bang cosmology imposes
stringent bounds on the masses of these right-handed neutrinos. We
show that in the framework of standard cosmology their predicted
abundance would violate the cosmological energy density bound.
This inconsistency between the alternative 422 model and the
standard cosmology implicitly assumes that the reheating
temperature is much higher than the freeze-out temperature of the
right-handed neutrinos, so that the right-handed neutrinos are
fully populated during the radiation-dominated era. However, if
the reheating temperature is low, $\sim$ MeV (not excluded by BBN
or any other observations), then the right-handed neutrino
production is highly suppressed. As a result there is a
significant range of parameters where the model can be reconciled
with cosmology. We conclude that low-scale quark-lepton
unification is a viable candidate for at least part of the new
physics suggested by the neutrino physics anomalies.

\vskip 0.4cm \noindent {\bf Acknowledgements} \vskip 0.4cm
\noindent T. L. Yoon acknowledges the support from OPRS and MRS.
He also likes to thank Chun Liu of Inst. Theor. Phys., Beijing,
for bringing the idea of the low reheating temperature scenario to
his attention.


\end{document}